\documentclass[usenatbib,usegraphicx,useAMS,usenatbib]{mn2e}

\newcommand{\uchii}{{UC H{\scriptsize II} }}
\newcommand{\hii}{{H{\scriptsize II} }}

\newcommand{\co}{$^{13}$CO}
\newcommand{\hco}{HCO$^+$}
\newcommand{\nh}{N$_2$H$^+$}
\newcommand{\chcn}{CH$_3$CN}
\newcommand{\choh}{CH$_3$OH}

\title[Mopra Observations of G305.2+0.2]{Mopra Observations
of G305.2+0.2: Massive Star Formation at Different Evolutionary Stages?}
\author[A. J. Walsh and M. G. Burton]{A. J. Walsh$^{1}$\thanks{E-mail:
awalsh@unsw.edu.au} and M. G. Burton$^{1}$\\
$^{1}$School of Physics, University of New South Wales, Sydney, NSW 2052, Australia}
\begin{document}



\maketitle

\label{firstpage}

\begin{abstract}
We have successfully used a new on-the-fly mapping technique with the Mopra
radiotelescope to image G305.2+0.2 in transitions of
\co , \hco , \nh , \chcn~and \choh. All these
species appear to be concentrated towards the infrared-quiet methanol
maser site G305A (G305.21+0.21). We suggest that this region contains an extremely deeply
embedded site of massive star formation, with comparable qualities
to the low mass Class 0 stage. The infrared-bright methanol maser site G305B (G305.21+0.20)
also exhibits emission in all the mapped transitions, but always at a lower level.
We suggest this is because it harbours a site of massive star formation older
and more developed than G305.21+0.21. All transitions appear to be extended beyond
the size of the Mopra beam (30\arcsec). \co~and \hco~line wings are suggestive of
an outflow in the region, but the spatial resolution of these data is insufficient
to identify the powering source. A narrow-lined (1.6\,km\,s$^{-1}$ compared to a typical line FWHM of
6.4\,km\,s$^{-1}$)  \nh~source (G305SW) is found
90\arcsec~to the south-west of the main star forming centres, which does not
correspond to any \chcn~or \choh~source, nor does it correspond well to
\co~or \hco~emission in the vicinity. We suggest this may be a massive, cold,
quiescent and possibly prestellar core.
\end{abstract}


\section{Introduction}
The study of the early stages of massive star formation (MSF) in our Galaxy has, until recently,
studied relatively litte. However, with the advent of telescopes,
particularly in the millimetre (mm) and sub-mm part of the electromagnetic
spectrum, it has become possible to study the earliest stages of MSF.
Here is the strongest part of their spectral energy distribution
(SED) measureable from the ground, as well as a myriad of
complex molecular transitions, allowing investigation of the
chemistry surrounding these young sources.

\begin{figure}
\includegraphics[width=\columnwidth]{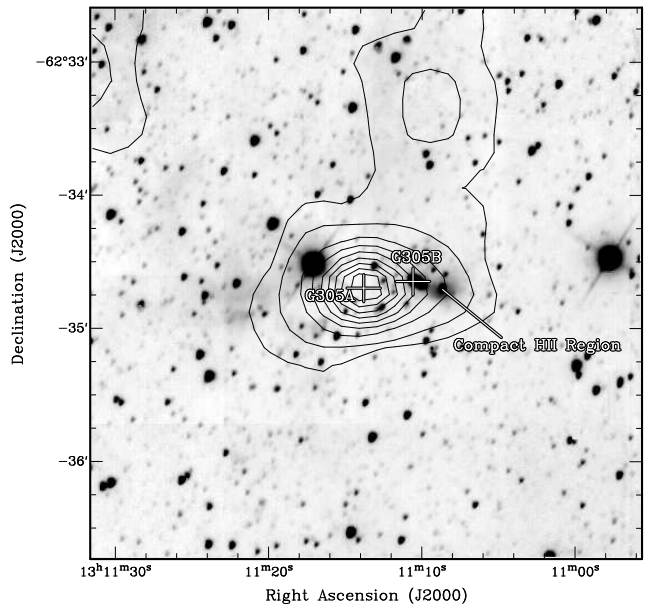}
\caption{Near-infrared K band (2.2\,$\mu$m) image of G305.2+0.2 \citep{walsh99}.
The two plus symbols represent the positions of the maser sites associated
with G305A and G305B. Contours are 1.2\,mm continuum emission with levels
15, 20, 30, 40, 50 60, 70, 80, 90\% of the peak 3.1\,Jy\,beam$^{-1}$ \citep{hill05}.}
\label{ksimb}
\end{figure}

G305.2+0.2 is a site of MSF in the southern Galactic plane shown in Figure
\ref{ksimb}. Class II methanol
masers were reported in two positions by \citet{norris93}:
G305.21+0.21 (hereafter G305A) and G305.20+0.21 (hereafter G305B), separated by
approximately 22\arcsec.
In G305A, the masers are extended along a line with
four individual maser spots aligned in
approximately the NE--SW direction \citep{norris93}. The maser flux is over
250\,Jy. G305B, on the other hand, contains two weak maser spots.
\citet{walsh99} observed this region in the near-infrared (see Figure \ref{ksimb})
and found that whilst
G305A is not associated with any infrared source, G305B is coincident with a
bright and very reddened infrared source. More sensitive near-infrared observations
by \citet{debuizer03} indicate that there is indeed a weak infrared source
coincident with G305A, however, it is not clear if it is
associated with the maser site or is instaead an unrelated foreground star.
Observations in the mid-infrared (10.5
and 20\,$\mu$m) by \citet{walsh01} confirm the infrared source associated with
G305B has a steeply rising spectral energy distribution towards longer wavelengths.
However, no mid-infrared source was found associated with G305A. Neither maser
site is coincident with any detected radio continuum source \citep{phillips98}.
However, one might be expected from G305B, since extrapolation of its SED
\citep{walsh01} indicates it is powered by a star bright enough to produce
an observable ultracompact (UC) \hii region.

The nature of this region is puzzling. It is not clear why such a bright
infrared source associated with G305B has no radio continuum counterpart.
Furthermore, it is not known why the brightest maser site, G305A is not associated
with a bright infrared source, nor a radio continuum source. Figure
\ref{ksimb} shows 1.2\,mm continuum contours, which show a source centred on
G305A, suggesting there is a dense dust core.
The emission appears to be slightly elongated in the direction of
G305B. Therefore it appears G305B may be associated with a weaker dust core,
which is spatially unresolved with these observations. In this paper,
we attempt to better understand these sources by looking for mm molecular line
emission, to trace the dense gas and chemistry of the region.

\section{Mopra Observations}

The Australia Telescope National Facility Mopra telescope is a 22\,m antenna
located 26\,km outside the town of Coonabarrabran in New South Wales, Australia.
It is at an elevation of 850 metres above sea level and at a latitude of
31$^\circ$ south.

The receiver comprises of a set of cryogenically cooled low noise SIS mixers
which operate between 85 and 115\,GHz The backend is a digital autocorrelator
which was used to provide a bandwidth of 64\,MHz over 1024 channels.

Observations were performed between 15-19 October, 2004. They typically took place
during overcast weather conditions.
Maps were obtained using a new on-the-fly (OTF) method for Mopra,
where the telescope scanned in the Right Ascension direction whilst data
were being accumulated. Data were reduced using livedata and gridzilla, which are
both AIPS++ packages written for the Parkes radiotelescope and adapted for
Mopra.

Table \ref{tab1} summarises the molecular and ionic line transitions mapped
towards G305.2+0.2.

\begin{table}
\caption{\protect\footnotesize{Properties of observed transitions}}
\label{tab1}
\begin{center}
\begin{tabular}{lcc}
\hline
Observed & Rest Frequency & Map\\
Line & (GHz) & Size\\
\hline
$^{13}$CO\,(1--0) & 110.201353 & 5\arcmin $\times$ 5\arcmin\\
HCO$^+$\,(1--0) & 89.188518 & 5\arcmin $\times$ 5\arcmin\\
N$_2$H$^+$\,(1--0) & 93.173258 & 5\arcmin $\times$ 5\arcmin\\
CH$_3$CN\,(5--4) & 91.987089 & 3.5\arcmin $\times$ 4\arcmin\\
CH$_3$OH\,($2_0$--$1_0$)A$^+$ & 96.741420 & 5\arcmin $\times$ 4\arcmin\\
CH$_3$OH\,($7_2$--$6_3$)A$^-$ & 86.615578 & 2.5\arcmin $\times$ 2\arcmin\\
\hline
\end{tabular}
\end{center}
\end{table}

\section{Results}

\begin{figure*}
\begin{tabular}{cc}
\includegraphics[width=0.8\columnwidth]{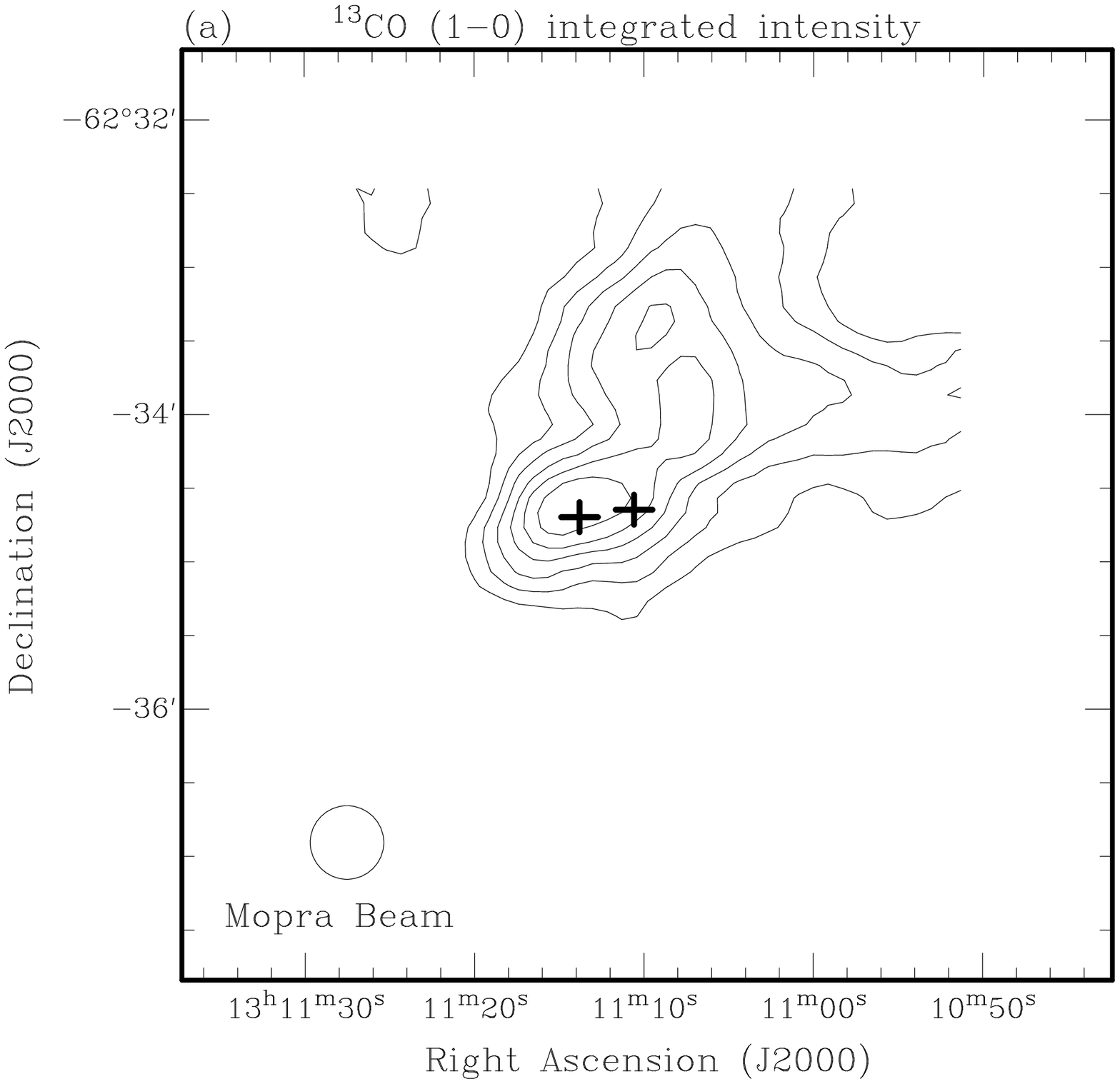}&\includegraphics[width=0.8\columnwidth]{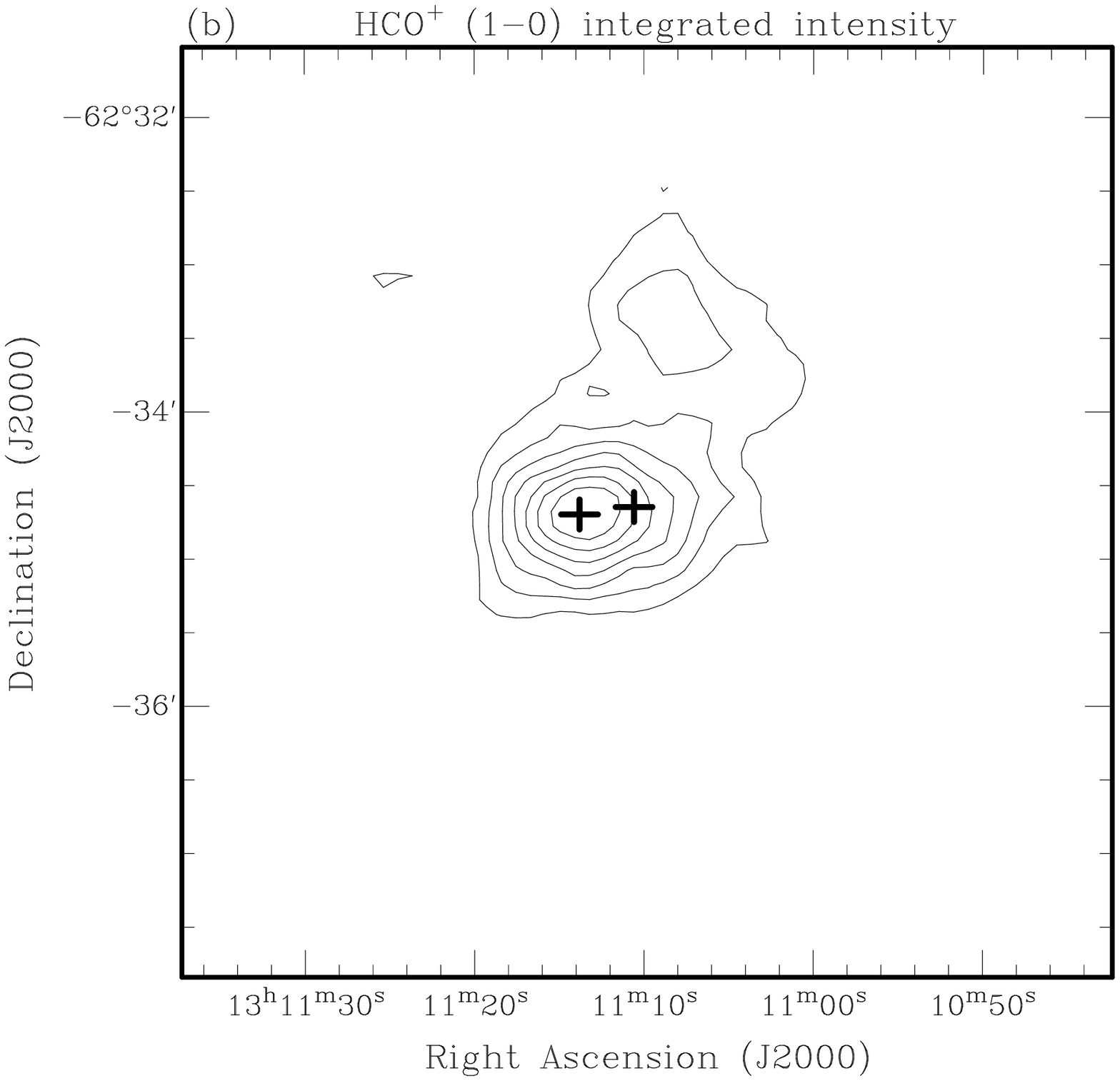}\\
\includegraphics[width=0.8\columnwidth]{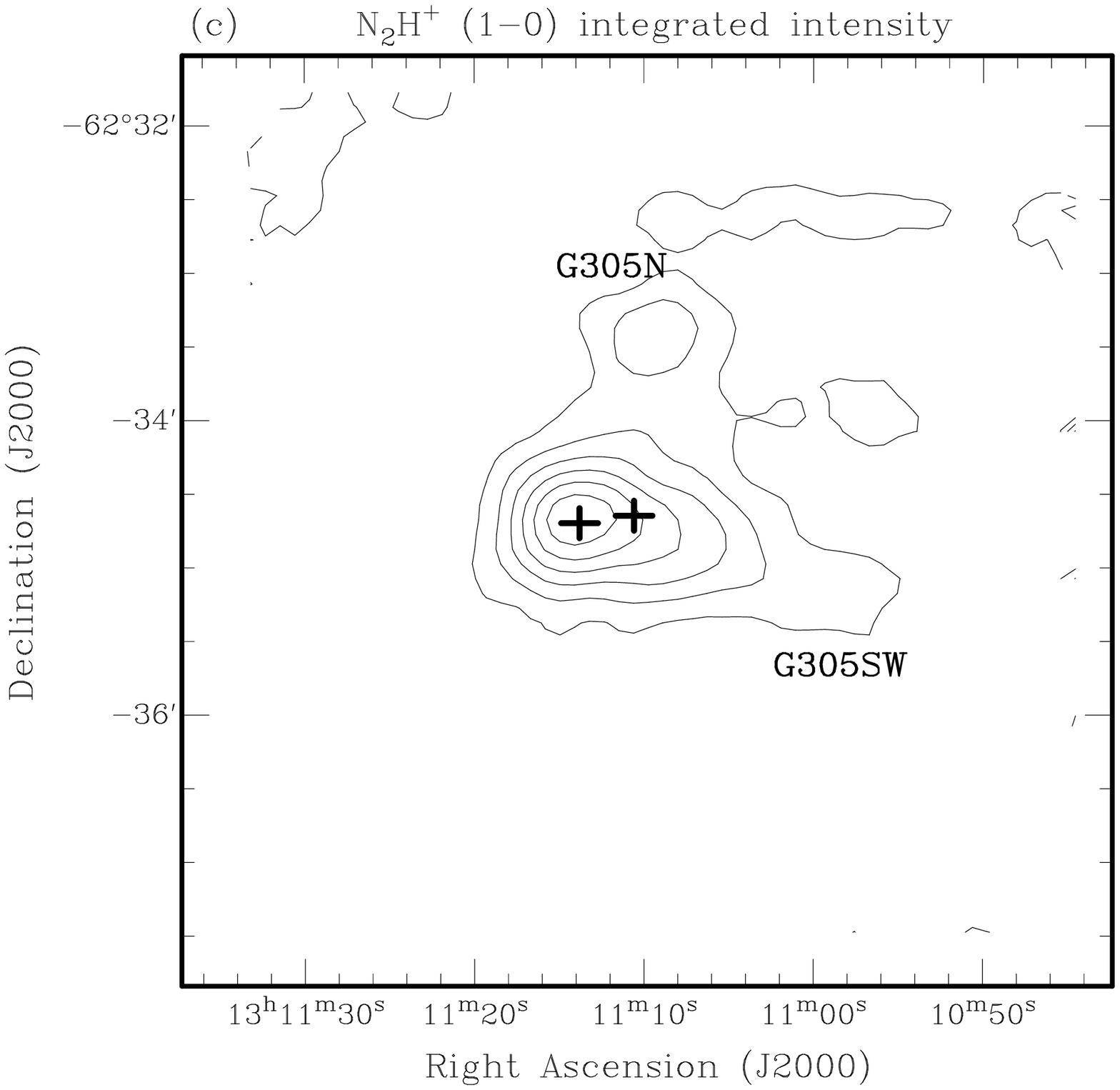}&\includegraphics[width=0.8\columnwidth]{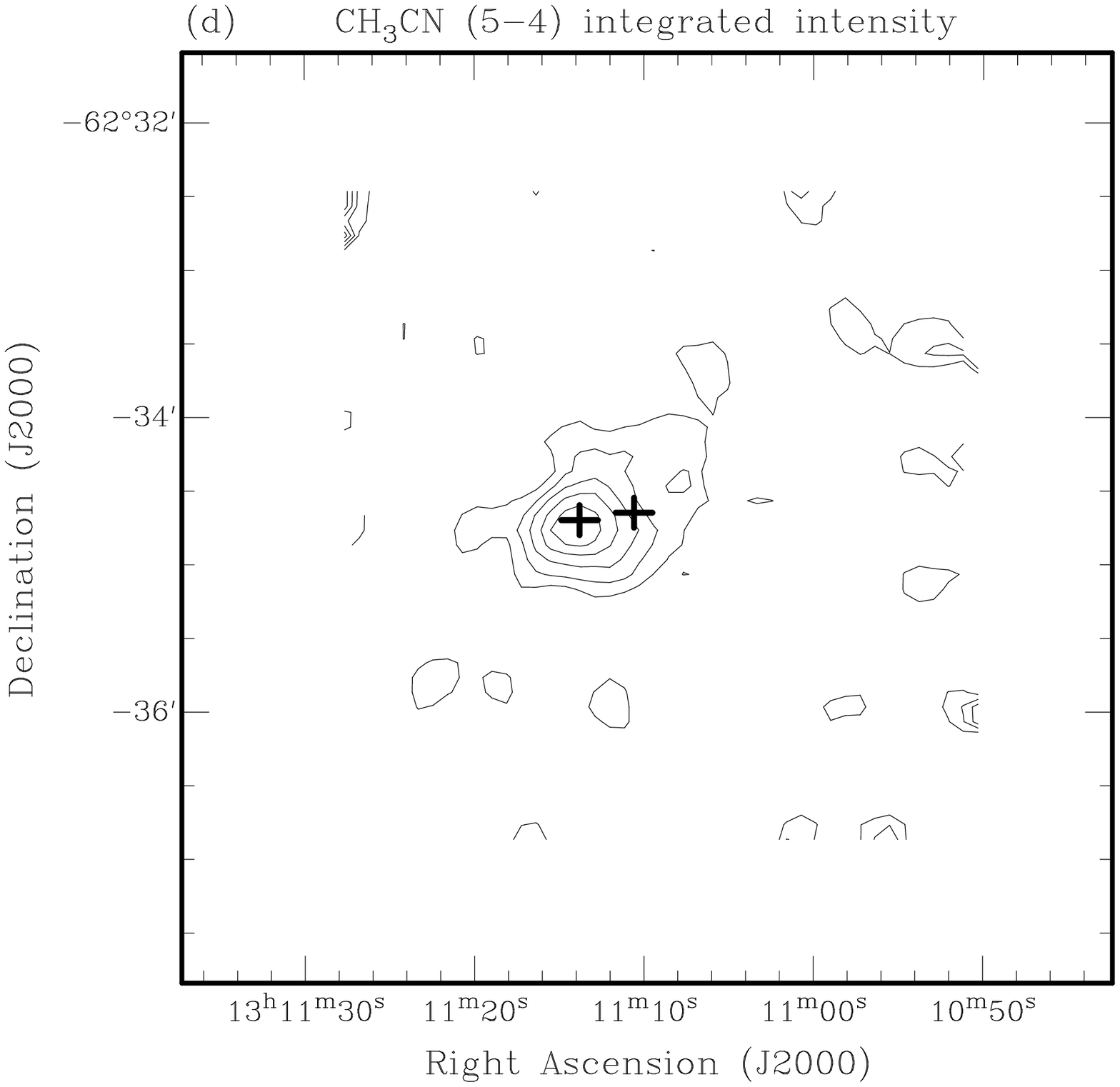}\\
\includegraphics[width=0.8\columnwidth]{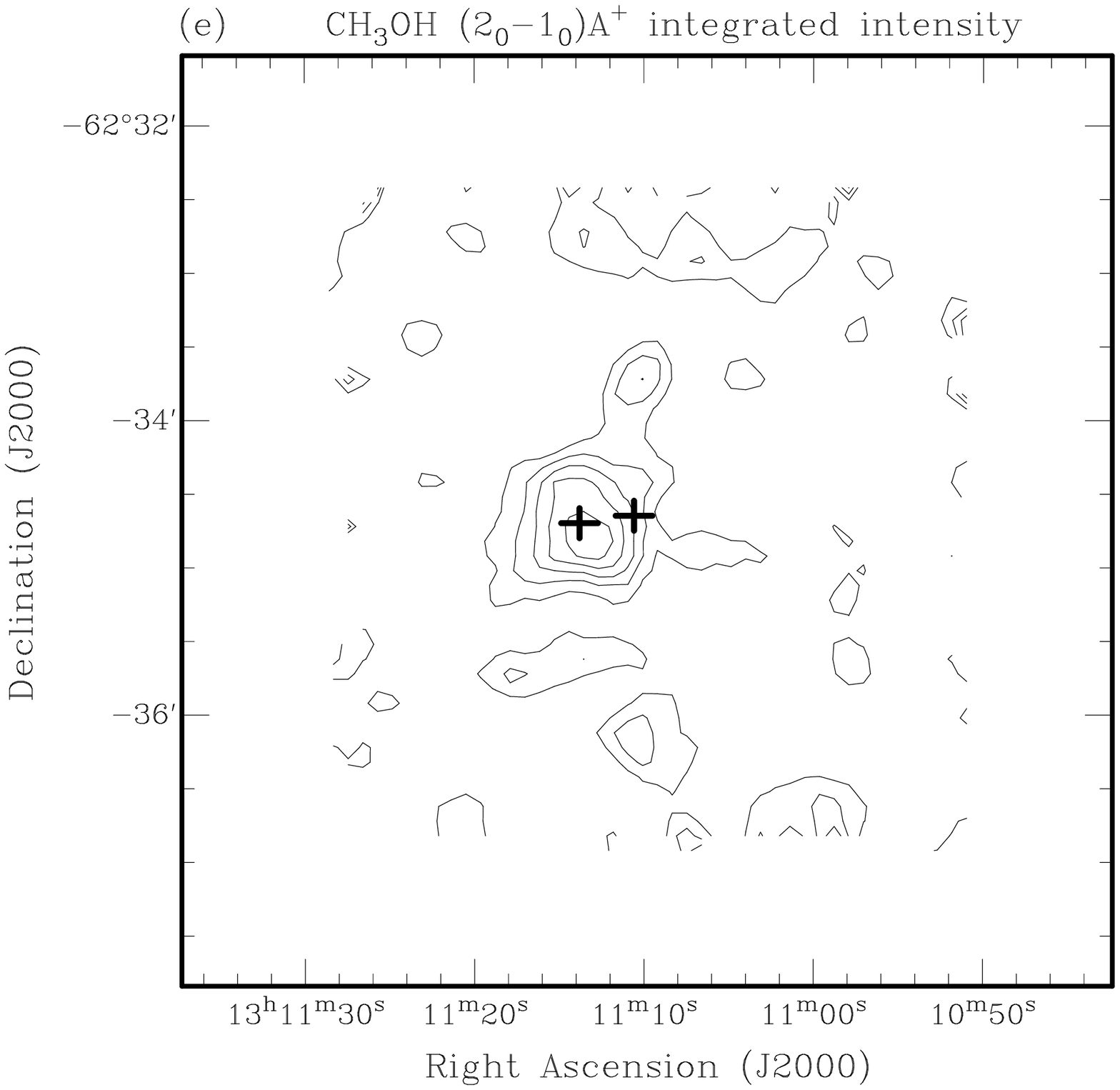}&\includegraphics[width=0.8\columnwidth]{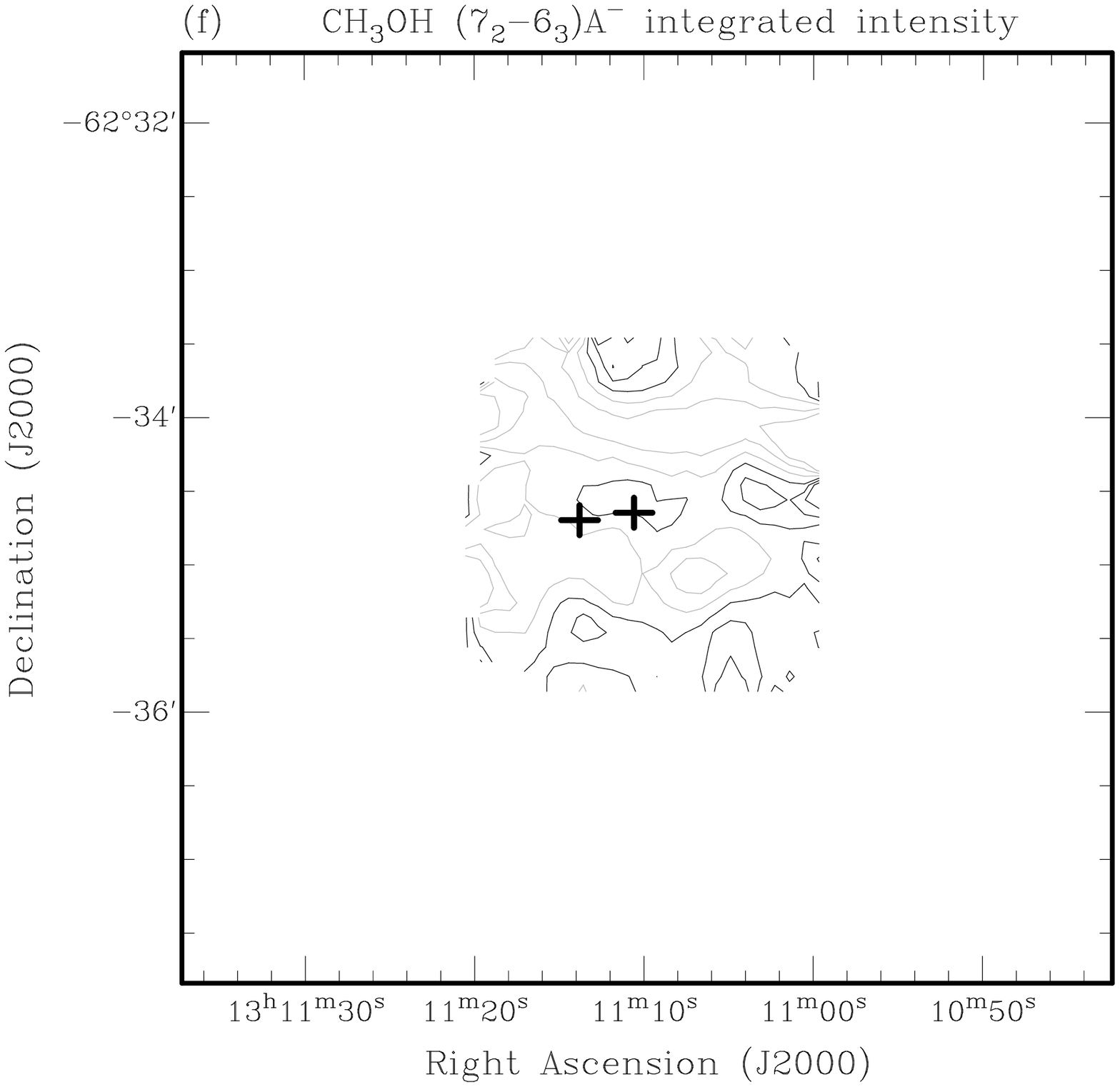}\\
\end{tabular}
\caption{Integrated intensity maps. Plus symbols show the positions of the maser sites
G305A and G305B.
{\bf (a)} $^{13}$CO (1--0): contour levels are 40, 50, ... 90\% of the peak, 43\,K\,km\,s$^{-1}$,
{\bf (b)} HCO$^+$ (1--0): contour levels are 30, 40, ... 90\% of the peak, 32\,K\,km\,s$^{-1}$,
{\bf (c)} N$_2$H$^+$ (1--0): contour levels are 40, 50, ... 90\% of the peak, 22\,K\,km\,s$^{-1}$,
{\bf (d)} CH$_3$CN (5--4): contour levels are 50, 60, ... 90\% of the peak, 4.6\,K\,km\,s$^{-1}$,
{\bf (e)} CH$_3$OH (2$_0$--1$_0$)A$^+$: contour levels are 50, 60, ... 90\% of the peak, 12\,K\,km\,s$^{-1}$,
{\bf (f)} CH$_3$OH (7$_2$--6$_3$)A$^-$: contour levels are 70, 80, 90\% of the peak, 2.4\,K\,km\,s$^{-1}$.}
\label{maps}
\end{figure*}

Integrated intensity maps of the observed transitions are shown
in Figure \ref{maps}. It can be seen from the figure that the five detected lines
appear to peak at the position of G305A. Only
CH$_3$OH\,($7_2$--$6_3$)A$^-$ transition at 86.6GHz was not detected.
Furthermore, $^{13}$CO, HCO$^+$, N$_2$H$^+$
and CH$_3$OH all show extended emission, with most of the emission centred on
G305A together with a spur of emission to the north extending approximately
2\arcmin~(hereafter G305N).
In addition to this, N$_2$H$^+$ also shows a spur to the SW of G305A,
extending about 2\arcmin~(hereafter G305SW). CH$_3$CN appears to show some extended emission
leading away from G305A to the west, but this lies close to the
orientation of the OTF scan and is close to the noise level in the map.
It may be an artifact. Therefore we consider
the CH$_3$CN emission to be essentially unresolved and centred on G305A, although
we do not discount the possibility
that there is extended emission to the west, close to the noise level of our
observations. Whilst there does appear to be emission in all
detected species at the position of G305B, it is always weaker than that found
coincident with G305A.

\begin{figure}
\includegraphics[width=\columnwidth]{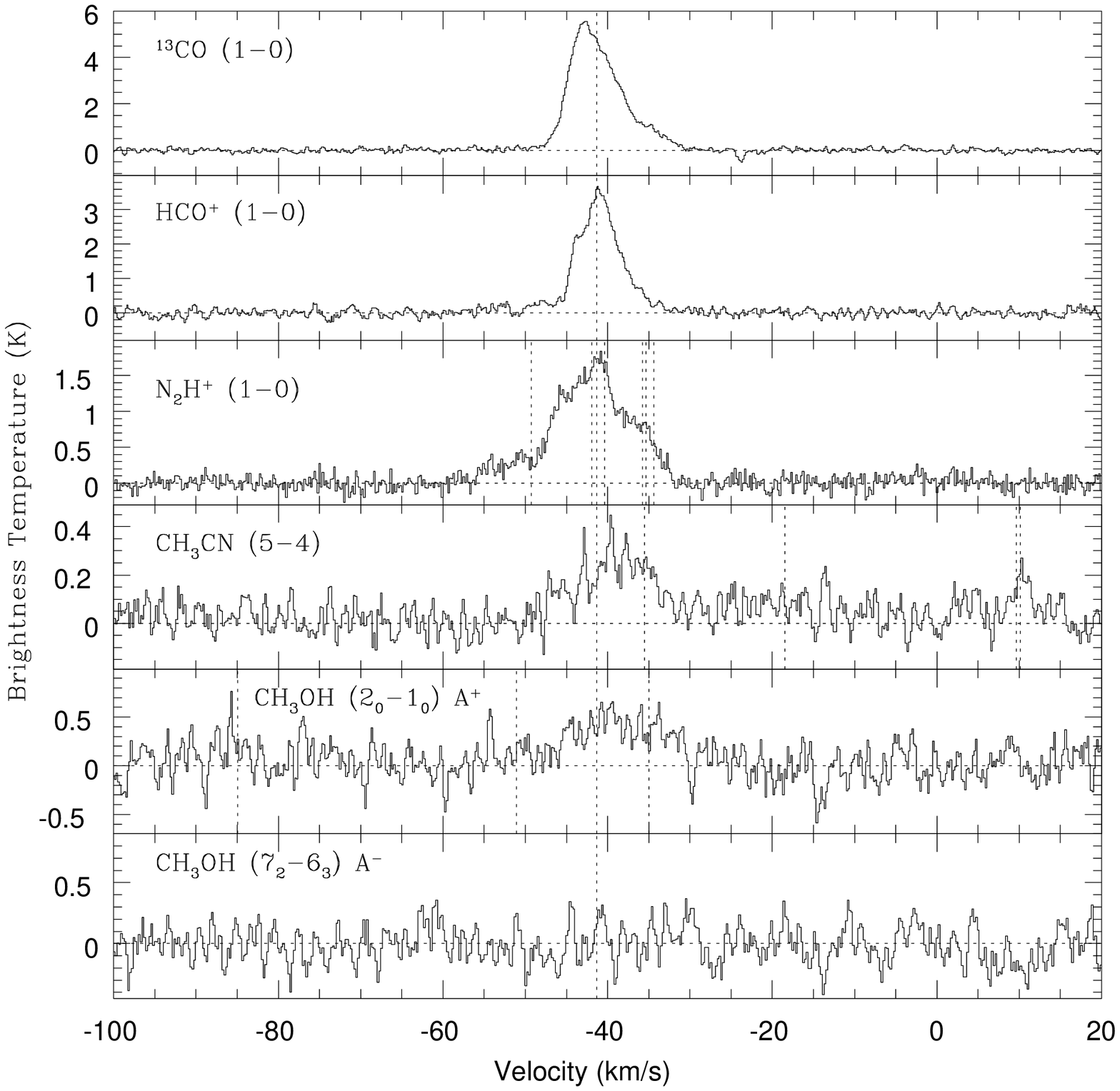}
\caption{Integrated spectra centred on G305A, and covering an area approximately
30\arcsec~(one beam). Dashed vertical lines indicate the assumed
${\rm V}_{\rm LSR}$ = -41.3\,km\,s$^{-1}$, which is determined from the peak of the HCO$^+$
spectrum. Multiple dashed lines indicate the positions of hyperfine components for
\nh~and rotational lines for \choh~and \chcn.}
\label{spec}
\end{figure}

Figure \ref{spec} shows integrated spectra of each line, covering
a square box of 30\arcsec~on a side (approximately one beam), centred on G305A,
which is the peak of the emission in most cases. $^{13}$CO, HCO$^+$ and N$_2$H$^+$
are all well detected at this position, whereas the spectra for both CH$_3$OH and
the CH$_3$CN transition do not clearly show a detection. This is partly due to
the poor baselines of the spectra, which are a result of the mediocre weather. However, we
can be sure that we have detected the CH$_3$OH\,($2_0$--$1_0$)A$^+$ and CH$_3$CN
(5--4) lines, as they clearly show up as 7 and 5$\sigma$ detections in Figure \ref{maps},
respectively. We note that both \co~and \hco~show some evidence for redshifted outflow wings.

\subsection{Column densities, abundances, LTE and virial masses}
For the three strongly detected species ($^{13}$CO, HCO$^+$ and N$_2$H$^+$), we can derive
physical parameters for the gas, on the assumption of optically thin emission.
Since N$_2$H$^+$ has hyperfine structure, we are
able to determine an optical depth at the positions of G305A, G305N and G305SW.
We have used the {\em hfs} hyperfine line fitting routine in {\sc class} \citep{forveille89}
to determine the optical depth of the N$_2$H$^+$ spectrum at the position of
G305A. We find the optical depth is $0.1 \pm 0.04$ in the strongest hyperfine component
(${\rm F}_1$F = 23 - 12). This is comparable to N$_2$H$^+$
optical depths found for a selection of massive molecular cloud cores \citep{pirogov03},
where the average optical depth is 0.5. We caution that the N$_2$H$^+$ spectrum is
heavily blended and may well show non-gaussian components. This may affect the optical
depth determination, and so we consider the formal error on the optical depth given above
to be optimistically small. However, we believe the error will not be so large that the
N$_2$H$^+$ emission is optically thick. At the positions of G305N and G305SW, we find
similar optical depths (0.1) of the strongest component. Therefore, we assume that
the N$_2$H$^+$ emission is optically thin throughout the field of view.

We derive column densities for each of the three species using equations \ref{eq2} and \ref{eq3}:

\begin{equation}
{\rm N} = \frac{3 \epsilon_0\,{\rm k}}{\pi^2 \nu \mu^2} \left(1-{\rm e}^{\frac{-{\rm h} \nu}{\rm k T}}\right)^{-1} \left(\frac{{\rm T}}{({\rm J}[{\rm T}]-{\rm J}[2.73])}\right) \int{\rm T}({\rm v})\;{\rm dv}
\label{eq2}
\end{equation}
where
\begin{equation}
{\rm J}[{\rm T}] = \frac{{\rm h} \nu}{{\rm k} \left({\rm e}^{\frac{{\rm h} \nu}{{\rm k} {\rm T}}} - 1\right)}
\label{eq3}
\end{equation}
where $\epsilon_0$ is the permittivity of free space, $\mu$ is the dipole moment (0.11\,Debye for $^{13}$CO,
3.3 Debye for HCO$^+$ and 3.4 Debye for N$_2$H$^+$, with 1 Debye = $1.13 \times 10^{-29}$\,C\,m) and
$\int{\rm T}({\rm v})\;{\rm dv}$ is the line integrated intensity. We assume the temperature of the gas
(T) to be 20\,K. The derived column densities are shown in Table \ref{tab2}.

\begin{table*}
\begin{minipage}{145mm}
\caption{\protect\footnotesize{Derived physical parameters for three cores.}}
\label{tab2}
\begin{center}
\renewcommand{\thefootnote}{\mbox{{$\alph{footnote}$}}}
\begin{tabular}{lcccccccc}
\hline
Core & Dust Derived & Integrated & Line & Radius & Column & Relative & M$_{\rm LTE}$ & M$_{\rm VIR}$\\
Name & Mass &Intensity & FWHM & (pc) & Density & Abundance & (M$_\odot$) & (M$_\odot$)\footnotemark[2]\\
& (M$_\odot$)\footnotemark[1] & (K\,\,km\,s$^{-1}$) & (km\,s$^{-1}$) && (${\rm cm}^{-2}$) &&&\\
\hline
\hline
\multicolumn{9}{c}{$^{13}$CO (1--0)} \\
\hline
G305A & 2700 & 110 & 6.8 & 0.7 & $1.4\times10^{17}$ & $2\times 10^{-6}$\footnotemark[3] & 1600 & 6800\\
G305N & 370 & 99 & 5.9 & 0.7 & $1.3\times10^{17}$ & $2\times 10^{-6}$ & 1400 & 5100\\
G305SW & $<$30\footnotemark[4] & 45 & 4.5 & $<$0.6 & $5.9\times10^{16}$ & $2\times 10^{-6}$ & 650 & $<$2400\\
\hline
\multicolumn{9}{c}{HCO$^+$ (1--0)}\\
\hline
G305A & 2700 & 63 & 7.3 & $<$0.6 & $1.4\times10^{14}$ & $3\times 10^{-9}$\footnotemark[5] & 990 & $<$9100\\
G305N & 370 & 28 & 5.1 & $<$0.6 & $6.0\times10^{13}$ & $3\times 10^{-9}$ & 440 & $<$3100\\
G305SW & $<$30 & 12 & 3.3 & $<$0.6 & $2.5\times10^{13}$ & $3\times 10^{-9}$ & 180 & $<$1200\\
\hline
\multicolumn{9}{c}{N$_2$H$^+$ (1--0)}\\
\hline
G305A & 2700 & 44 & 6.9 & $<$0.6 & $8.1\times10^{13}$ & $5\times 10^{-10}$\footnotemark[6] & 3600 & $<$8200\\
G305N & 370 & 19 & 5.2 & $<$0.6 & $3.5\times10^{13}$ & $5\times 10^{-10}$ & 1500 & $<$3200\\
G305SW & $<$30 & 13 & 1.5 & $<$0.6 & $2.4\times10^{13}$ & $5\times 10^{-10}$ & 1100 & $<$540\\
\hline
\end{tabular}
\end{center}
\begin{flushleft}
$^a$ \citet{hill05}
$^b$ Virial masses are calculated assuming a uniform density sphere. Virial mass upper
limits are quoted in some cases because the emission is spatially unresolved.
$^c$ \citet{dickman78}
$^d$ Non-detection of dust continuum emission. The 1$\sigma$ upper limit is given.
$^e$ \citet{bergin97}
$^f$ \citet{pirogov03}
\end{flushleft}
\end{minipage}
\end{table*}

Table \ref{tab2} also quotes core masses derived from dust continuum measurements
\citep{hill05} for both G305A and G305N. \citet{hill05} did not detect any dust continuum
emission associated with G305SW, and so we quote an upper limit based on the rms
noise level.

Assuming LTE, we estimate the total gas mass, based on the column density of each line.
In order to do so, we need to know the abundance of each species relative to
hydrogen. We have chosen representative examples of relative abundances for $^{13}$CO,
HCO$^+$ and N$_2$H$^+$ from the literature, as given in Table \ref{tab2}. We caution that
the LTE derived mass is heavily dependant on the relative abundance of each species, and
discuss this further in \S4.

We can also derive virial masses according to equation \ref{eq1} \citep{pirogov03}:
\begin{equation}
\left(\frac{{\rm M}_{\rm vir}}{{\rm M}_\odot}\right) = 210 \left(\frac{\rm r}{\rm pc}\right) \left(\frac{\Delta{\rm v}}{{\rm km\,s}^{-1}}\right)^2
\label{eq1}
\end{equation}
where r is the radius of the core and $\Delta{\rm v}$ is the line FWHM. We assume the virial
mass is determined for a sphere of uniform density. For a sphere with a density
profile that falls off as the square of the radius, then the virial mass is reduced
by a factor of 1.68. Values for the virial mass are also shown in Table \ref{tab2}.

\section{Discussion}

\subsection{G305A}
\citet{hill05} calculate a core mass of $2.7 \times 10^3\, {\rm M}_\odot$ for the 1.2\,mm
continuum peak centred on G305A, assuming a distance of 3.9\,kpc. Compared to
other cores in their survey, \citet{hill05} find that this mass is
typical for cores that form massive stars. Therefore, even though no observed
source has been detected, we believe that G305A harbours enough mass to form a massive star.
Furthermore, the Class II methanol maser site associated with G305A strongly suggests there is
MSF taking place here because no other Class II methanol maser site has been conclusively
found associated with anything other than MSF, despite previous searches \citep{phillips98b,minier03}.

Together with the presence of methanol masers and 1.2\,mm continuum emission, shown in Figure \ref{ksimb},
the new Mopra observations of G305A conclusively show that the core is associated
with bright mm-line emission. Since there is no known infrared
or radio continuum tracer, G305A is likely to be an extremely young site of MSF.
It compares well to the Class 0 phase of low mass star formation, as it shows strong mm
continuum emission, but no detectable IR emission. The Class 0 phase of low mass star formation
is characterised by these observational constraints, but is also considered to be a phase
of active accretion \citep{andre93}. We caution that we currently have no evidence for ongoing
accretion in this high mass case, apart from the indirect evidence of the red outflow wings, as
seen in Figure \ref{spec}. The spatial resolution of the Mopra data is insufficient to establish
whether or not G305A is the powering source of the outflow. We cannot assume accretion is taking place because
accretion processes are expected to proceed on much shorter timescales than for low mass star formation.
Therefore, it is conceivable that accretion has ceased in G305A but it still remains extremely deeply
embedded and is thus not seen in the IR. On the other hand, G305A may be in the
short-lived accretion phase. In any case, G305A is an excellent test case for accretion processes
in MSF.

Because G305A does not show any signs of near-infrared \citep{walsh99} or mid-infrared
\citep{walsh01} emission up to 20\,$\mu$m, the presence of strong Class II methanol
maser emission is somewhat of a mystery. Class II methanol masers are thought to be
pumped by infrared photons from warm (T$\geq$ 150\,K)
dust \citep{cragg02}. Current observations
extend only to 20\,$\mu$m, and so the pumping photons may well be found only at longer
wavelengths. However, the peak of a 150\,K black body should be about 20$\mu$m, suggesting
that there should be a bright mid-infrared source where there is bright methanol maser
emission. In any case, G305A should be a useful testbed of methanol maser pumping
regimes.

Table \ref{tab2} shows that the LTE and virial derived masses from $^{13}$CO, HCO$^+$ and
N$_2$H$^+$ are all consistent within a factor of a few. LTE masses derived using
$^{13}$CO and HCO$^+$ appear lower than the mass derived from the dust measurements, however it is
difficult to pinpoint the cause of this, since there are many possible reasons, for instance
the lines may be tracing slightly cooler gas and so the assumption of 20\,K gas temperature
may be incorrect. We note that if the gas temperature were only 10\,K, then the LTE masses
would increase to 4000M$_\odot$ for $^{13}$CO and 2500M$_\odot$ for HCO$^+$. Another
possibility is that there is some depletion of these species onto dust grains, causing an
apparent drop in the LTE mass. It is also possible that the emission is optically thick,
which would also have the effect of reducing the apparent LTE mass. At this stage, it is
impossible for us to disentangle these contributing factors.

\subsection{G305B}

G305B is coincident with a deeply embedded infrared source with a steeply rising SED.
At 10\,$\mu$m, this is the brightest source within the field of view. Based on the
infrared data, \citet{walsh01} expected that such a bright IR source should produce
a detectable radio continuum source. However, none was detected. In fact,
\citet{walsh01} show G305B to have one of the largest discrepancies between
the bright infrared emission and the undetected radio continuum emission in their sample
of 31 sources.
This suggests that G305B is also at a very early stage of evolution. Presumably, a
massive star has already formed at the centre due to the presence of the near-infrared
source, yet no detectable radio continuum source has developed. It is unclear exactly
why this might happen, although theories such as accretion through an \uchii region \citep{keto03}
could explain this.

\subsection{G305SW}

\begin{figure}
\includegraphics[width=\columnwidth]{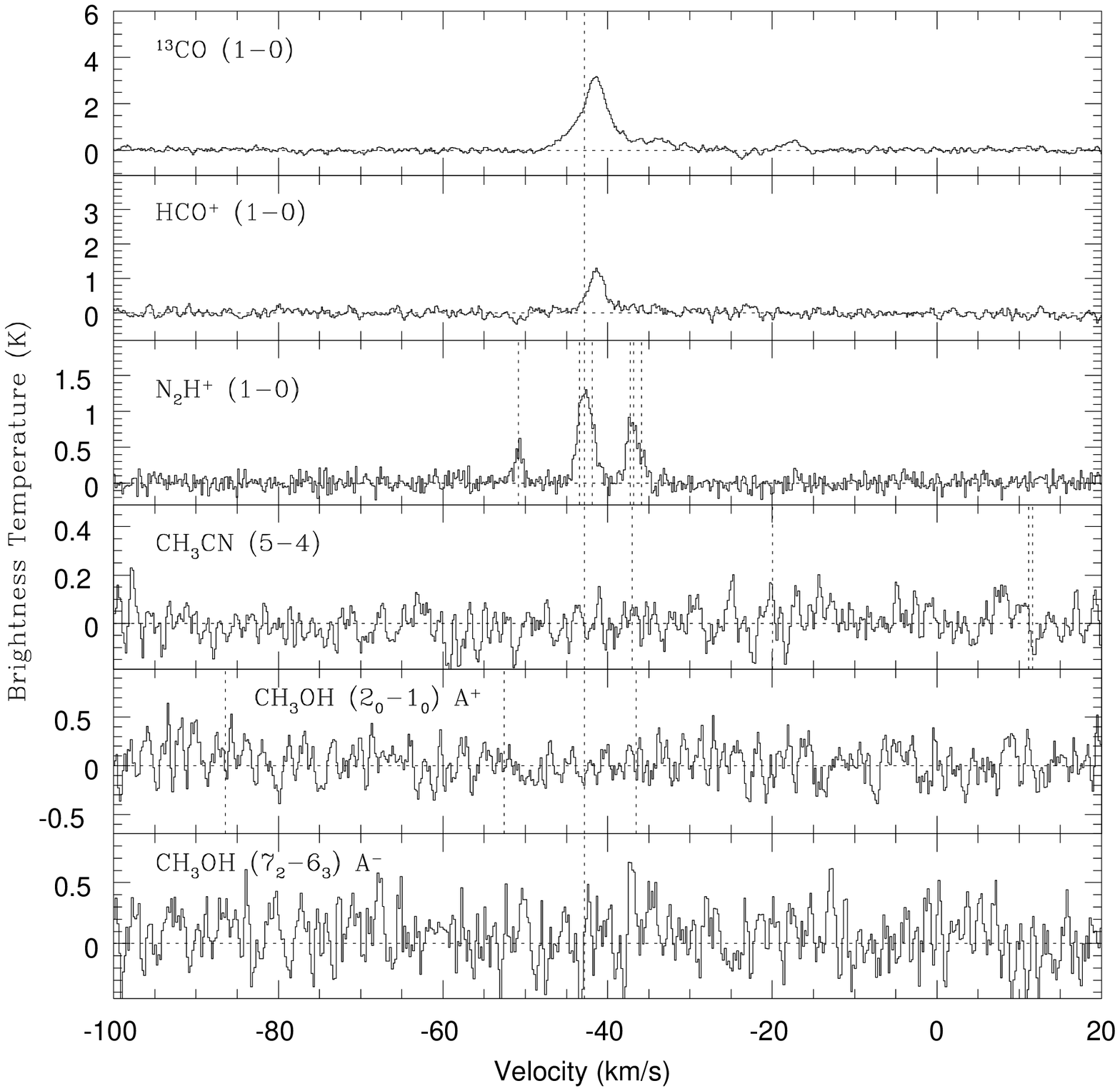}
\caption{Integrated spectra centred on G305SW, and covering an area approximately
30\arcsec~(one beam). Dashed vertical lines indicate the assumed
${\rm V}_{\rm LSR}$ = $-$42.8\,km\,s$^{-1}$, which is determined from a hyperfine
fit to the N$_2$H$^+$ spectrum. Multiple dashed lines indicate the positions of
hyperfine components for \nh~and rotational lines for \choh~and \chcn.}
\label{g305swspec}
\end{figure}

N$_2$H$^+$ emission is seen to extend to the south-west of the two maser sites,
ending in the peak G305SW. G305SW appears only in the N$_2$H$^+$
integrated intensity map. This is an unexpected result because the
critical density for N$_2$H$^+$ ($2 \times 10^5\, {\rm cm}^{-3}$; Ungerechts et al. 1997)
is two orders of magnitude higher than that
of $^{13}$CO ($1 \times 10^3\, {\rm cm}^{-3}$; Ungerechts et al. 1997).
Therefore, we would expect any high density core such as G305SW
to also exhibit emission in low density tracers. Figure \ref{g305swspec} shows
spectra at the position of G305SW, integrated over a square box 30\arcsec~on
a side. The figure shows that whilst there is some $^{13}$CO and HCO$^+$ emission
at this position, it is much weaker in comparison to G305A. Furthermore, the 
$^{13}$CO and HCO$^+$ emission does not match in radial velocity to the N$_2$H$^+$,
with a difference of 1.5\,km\,s$^{-1}$: the radial velocity of $^{13}$CO and HCO$^+$
more closely matches that of the emission associated with G305A.
With such a large velocity difference, it
is unlikely that this is the result of self-absorbed gas moving along the line of sight
(in this case the motions would be outward, rather than the usual infall motions).
We conclude that most of the low density gas, traced by $^{13}$CO and HCO$^+$, is not
associated with G305SW and is probably associated with G305A.

\begin{figure}
\includegraphics[width=0.8\columnwidth,angle=-90]{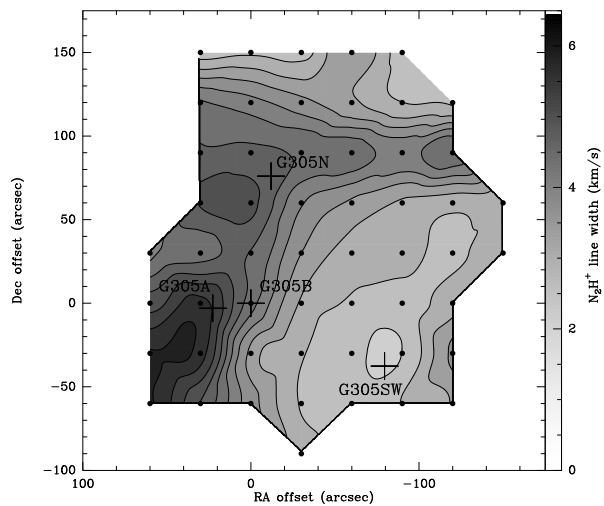}
\caption{N$_2$H$^+$ line width distribution. The lowest contour, encircling G305SW,
is at
0.5\,km\,s$^{-1}$. Contours increase in 0.5\,km\,s$^{-1}$ steps. Filled circles
indicate positions where the N$_2$H$^+$ line was strong enough to fit the hyperfine
components of the spectrum and reliably determine the line width. Plus symbols
indicate the positions of the main regions of interest. The map is centred on
G305B.}
\label{lwidth} 
\end{figure}

G305SW is also unusual in that the N$_2$H$^+$ line width is only
1.6\,km\,s$^{-1}$. This is very narrow compared to the line width 
of the N$_2$H$^+$ spectrum in Figure \ref{spec} for G305A of 6.4\,km\,s$^{-1}$. Figure
\ref{lwidth} shows the distribution of N$_2$H$^+$ line widths across the field of
view, using those spectra with strong enough emission to fit the hyperfine
structure. The minimum line width is clearly seen at a position coincident with G305SW.

Figure \ref{ksimb} shows that
no 1.2\,mm continuum emission was detected at the position of G305SW. Assuming a black body
temperature of 20\,K, a gas to dust ratio of 100:1 and a mass absorption coefficient of
0.1\,m$^2$\,kg$^{-1}$, any source located at G305SW must be less than 30\,${\rm M}_\odot$. This
is an unusually low mass compared to G305A. Even if our assumption of the dust temperature is incorrect,
a dust temperature of 10\,K yields a mass upper limit of only 75\,${\rm M}_\odot$.
This can be compared to the LTE-derived mass (Table \ref{tab2}) of 1100\,M$_\odot$. Thus,
there appears to be a discrepancy between the dust upper limit and the N$_2$H$^+$ source
by at least an order of magnitude.

The virial mass is then $< 540\,{\rm M}_\odot$. Equation \ref{eq1} assumes a uniform density profile.
However, if we assume a density profile that falls off as ${\rm r}^{-2}$ then the virial mass will be
$< 320\,{\rm M}_\odot$. In either case, the virial mass upper limit is about an order of magnitude higher than
the mass upper limit calculated from the 1.2\,mm continuum observations. Our virial mass estimation
is only an upper limit because the N$_2$H$^+$ emission is unresolved with the Mopra observations.
It is certainly conceivable that G305SW is much smaller than this, bringing the virial and dust masses
more in line, however the N$_2$H$^+$ LTE mass is still much larger than expected.

What is the nature of G305SW?
One possible explanation for the lack of other tracers than N$_2$H$^+$ is that the
region may be very cold and dense, where all other species have frozen out of
the gas phase onto grains, leaving N$_2$H$^+$ -- one of the last species to freeze
out \citep{tafalla02} -- in the gas phase. The fact that the 1.2\,mm continuum observations
indicate there is less than 100\,${\rm M}_\odot$ in G305SW and yet the \nh~emission is consistent
with an LTE mass about an order of magnitude larger, confirms that chemistry must play
a significant role. G305SW may well be a 
cold, quiescent core perhaps about to undergo star formation. The narrow N$_2$H$^+$
linewidth of G305SW also supports this scenario. The fact that the virial mass may be much larger
than the mass derived from the dust continuum measurements suggests that G305SW may well be an
unbound, transient phenomenon, and we cannot rule this out.

To fully assess the nature of G305SW, it is clear that higher spatial resolution observations are
needed. It is conceivable that the \nh~emission breaks up into a number of small cores at
higher resolution, in which case the individual cores may well be bound.

\subsection{G305N}
G305N appears as a peak of 1.2\,mm continuum emission, as well as in the $^{13}$CO,
HCO$^+$, N$_2$H$^+$ and CH$_3$OH ($2_0 -- 1_0$) A$^+$ integrated intensity maps,
but does not show
any infrared, radio continuum nor methanol maser emission. \citet{hill05} calculate a
core mass of $3.7 \times 10^2\,{\rm M}_\odot$.
\citet{hill05} detect eight 1.2\,mm continuum sources in their field of view around G305.2+0.2,
with masses ranging from 39\,${\rm M}_\odot$ to $2.7 \times 10^3\,{\rm M}_\odot$, of which
G305N is a typical source. They commonly find other 1.2\,mm continuum sources
with comparable masses to G305N, so it is possible that this source will form massive stars.
Since G305N does not show any signs of MSF -- no masers, radio
continuum or bright infrared sources -- it may also be a good candidate for examining
the early stages of MSF.

\section{Conclusions}

We have used the Mopra radiotelescope to image mm line transitions of \co , \hco , \nh ,
\chcn~and \choh~in the massive star forming region G3052.+0.2.
We show that the main centre of line emission is G305A, a methanol maser
site that has no detected infrared or radio continuum emission and is likely to
be a massive star in an early stage of formation, similar to the Class 0 stage
of low mass star formation. G305B is a second methanol maser site, coincident with a bright,
red infrared source whose bolometric luminosity suggests it should produce a detectable
\hii region, yet none has been detected. This suggests that G305B may also be at a very
early stage of evolution, but probably older than G305A.

G305SW is a newly detected source which only shows up clearly in \nh~emission. Its lack of
other tracers and narrow linewidth suggest that it may be a cold, quiescent
prestellar core, although it may also be an unbound, transient phenomenon.

Other prestellar cores may exist (eg. G305N), and on the other hand more evolved massive
star forming sites occur (eg. the compact H{\scriptsize II} region to the west of G305B).
Thus, this region shows a range of massive star formation sites at different evolutionary
stages within the GMC complex.



\label{lastpage}

\end{document}